\documentclass[10pt]{article}
\usepackage{amsfonts,xurl,hyperref,amsmath,breqn,fancyvrb,fvextra,customdice}
\usepackage[a4paper,margin=2cm]{geometry}
\begin{document}
\title{A note on Erdős's mysterious remark\footnote{Dedicated to Tomás Recio's 75th birthday}}
\author{Zoltán Kovács, PHDL Linz, \texttt{zoltan.kovacs@ph-linz.at}}
\maketitle
\abstract{We give an alternative proof of the statement, by using elimination from algebraic geometry, that the only set $S\subset\mathbb{R}^2$, $\left|S\right|=6$ such
that all subsets that form a triangle are isosceles triangles, is the regular pentagon with its center. Our proof can be extended to answer some related questions raised by Erdős.}

\section{The problem}
Paul Erdős at his last years published a list of unsolved problems, including some geometric questions, which were raised by him
during the years he worked very actively and extremely fruitfully in many parts of mathematics.

In this paper, we give a detailed proof for a remark of Erdős's open problem \#91 (see \url{https://www.erdosproblems.com/91})
which was already communicated by Erdős that it had been solved by ``a colleague'' \cite{Er90,Er97}, but the proof was
actually never published.
In \cite{Er93} he mentions a similar problem which has not yet been solved, and also he labels the question as ``probably not difficult''. It can be formalized in this way: \textit{Find a set of 6 points in the plane such that for each triplet the
triangle it defines is isosceles.} It is easy to see that a regular pentagon with its center is one solution.
To prove: there is no other solution.

In 1946, Erdős proposed the very similar Problem E~735 whose generalization for $\mathbb{R}^n$ is currently numbered as \#503 in the above-mentioned database:
\begin{quote}
Six points can be arranged in the plane so that all triangles formed by
triples of these points are isosceles. Show that seven points in the plane cannot be so arranged. What is the least number of points in space which cannot be so arranged?
\end{quote}
For $n=2$, this was solved by L.M.~Kelly and the solution was published the following year in \cite{ErKe47}. Kelly used case distinctions to study the situation, and finally he concluded that the first five points must form a regular pentagon and found the ``characterized plane isosceles 6-points as the vertices and center of a regular pentagon''. Kelly's proof actually addresses Erdős's remark, but Erdős was probably unaware, unsatisfied or not fully convinced with Kelly's geometric argumentation which took about 2 pages.

In the first part of this paper we show an alternative proof for E~735 for the plane. Then we address the remark in \#91 by extending our proof.

\section{Setting up the equations}
We set up a set of polynomial equations that describe the geometric situation algebraically.
Let us denote the elements of the searched set $S$ by $A_1=(x_1,y_1)$, $A_2=(x_2,y_2)$, $\ldots$, $A_6=(x_6,y_6)$.
Without loss of generality we can assume the $A_1=(-1,0)$ and $A_2=(1,0)$.
By using Pythagorean distances, it is possible to express that the triangle $A_iA_jA_k$ ($\{i,j,k\}\subset \{1,2,\ldots,6\}$)
is isosceles: the product of
$$\left((x_i-x_j)^2+(y_i-y_j)^2\right)-\left((x_j-x_k)^2+(y_j-y_k)^2\right),$$
$$\left((x_j-x_k)^2+(y_j-y_k)^2\right)-\left((x_k-x_i)^2+(y_k-y_i)^2\right)$$ and
$$\left((x_k-x_i)^2+(y_k-y_i)^2\right)-\left((x_i-x_j)^2+(y_i-y_j)^2\right)$$ equals to $0$.
For all $\binom{6}{3}=20$ triplets of $S$ we set up such a product (a degree 6 polynomial) and denote them
by $p_1,p_2,\ldots,p_{20}$.

To avoid degeneracy, we also assume that for each $\{i,j\}\subset \{1,2,\ldots,6\}$, $\{i,j\}\ne\{1,2\}$,
$A_i\neq A_j$, that is,
$$(x_i-x_j)^2+(y_i-y_j)^2\neq0.$$ The left hand side of these equations, $\binom{6}{2}-1=14$ polynomials
are denoted by $q_1,q_2,\ldots,q_{14}$. By using Rabinowitsch's trick \cite{Ra29}, we can express the non-degeneracy by considering
the polynomial
$$\prod_{i=1}^{14}q_i \cdot t-1$$ where $t$ is a dummy variable to express that none of the $q$'s can be $0$.
This polynomial is of degree 29.

Finally, we compute the elimination ideal
$$\left<p_1,p_2,\ldots,p_{20},\prod_{i=1}^{14}q_i\cdot t-1\right>\cap\mathbb{Q}[x_3,y_3]$$ which delivers
the union of all possible sets $S$ that satisfy the assumption.

Unfortunately, this problem is computationally difficult. It is not known whether any successful run has ever found
the solution. Therefore, we simplify the question to 5 points.

\section{The problem with 5 points}
It is clear that if a set $S$ with 6 points can be found, then any subset $S'\subset S$, $\left|S\right|<6$ will
also fulfill the requirements. Therefore, we drop all equations that contain the coordinates of $A_6$, and
there remain only $\binom{5}{3}=10$ $p$'s and $\binom{5}{2}-1=9$ $q$'s. Hopefully, the new elimination
ideal achieved is easier to compute. In addition, the remaining point $A_6$ will hopefully be a simple task to determine geometrically.

The following program, written in the Giac computer algebra system, successfully finds the union of all possible sets $S'$:

{\footnotesize
\begin{Verbatim}[breaklines=true,breakafter=*]
x1:=-1
y1:=0
x2:=1
y2:=0
p1:=((x3-x1)^2+(y3-y1)^2-(x3-x2)^2-(y3-y2)^2)*((x1-x2)^2+(y1-y2)^2-(x1-x3)^2-(y1-y3)^2)*((x2-x3)^2+(y2-y3)^2-(x2-x1)^2-(y2-y1)^2)
p2:=((x4-x1)^2+(y4-y1)^2-(x4-x2)^2-(y4-y2)^2)*((x1-x2)^2+(y1-y2)^2-(x1-x4)^2-(y1-y4)^2)*((x2-x4)^2+(y2-y4)^2-(x2-x1)^2-(y2-y1)^2)
p3:=((x5-x1)^2+(y5-y1)^2-(x5-x2)^2-(y5-y2)^2)*((x1-x2)^2+(y1-y2)^2-(x1-x5)^2-(y1-y5)^2)*((x2-x5)^2+(y2-y5)^2-(x2-x1)^2-(y2-y1)^2)
// p4:=((x6-x1)^2+(y6-y1)^2-(x6-x2)^2-(y6-y2)^2)*((x1-x2)^2+(y1-y2)^2-(x1-x6)^2-(y1-y6)^2)*((x2-x6)^2+(y2-y6)^2-(x2-x1)^2-(y2-y1)^2)
p5:=((x4-x1)^2+(y4-y1)^2-(x4-x3)^2-(y4-y3)^2)*((x1-x3)^2+(y1-y3)^2-(x1-x4)^2-(y1-y4)^2)*((x3-x4)^2+(y3-y4)^2-(x3-x1)^2-(y3-y1)^2)
p6:=((x5-x1)^2+(y5-y1)^2-(x5-x3)^2-(y5-y3)^2)*((x1-x3)^2+(y1-y3)^2-(x1-x5)^2-(y1-y5)^2)*((x3-x5)^2+(y3-y5)^2-(x3-x1)^2-(y3-y1)^2)
// p7:=((x6-x1)^2+(y6-y1)^2-(x6-x3)^2-(y6-y3)^2)*((x1-x3)^2+(y1-y3)^2-(x1-x6)^2-(y1-y6)^2)*((x3-x6)^2+(y3-y6)^2-(x3-x1)^2-(y3-y1)^2)
p8:=((x5-x1)^2+(y5-y1)^2-(x5-x4)^2-(y5-y4)^2)*((x1-x4)^2+(y1-y4)^2-(x1-x5)^2-(y1-y5)^2)*((x4-x5)^2+(y4-y5)^2-(x4-x1)^2-(y4-y1)^2)
// p9:=((x6-x1)^2+(y6-y1)^2-(x6-x4)^2-(y6-y4)^2)*((x1-x4)^2+(y1-y4)^2-(x1-x6)^2-(y1-y6)^2)*((x4-x6)^2+(y4-y6)^2-(x4-x1)^2-(y4-y1)^2)
// p10:=((x6-x1)^2+(y6-y1)^2-(x6-x5)^2-(y6-y5)^2)*((x1-x5)^2+(y1-y5)^2-(x1-x6)^2-(y1-y6)^2)*((x5-x6)^2+(y5-y6)^2-(x5-x1)^2-(y5-y1)^2)
p11:=((x4-x2)^2+(y4-y2)^2-(x4-x3)^2-(y4-y3)^2)*((x2-x3)^2+(y2-y3)^2-(x2-x4)^2-(y2-y4)^2)*((x3-x4)^2+(y3-y4)^2-(x3-x2)^2-(y3-y2)^2)
p12:=((x5-x2)^2+(y5-y2)^2-(x5-x3)^2-(y5-y3)^2)*((x2-x3)^2+(y2-y3)^2-(x2-x5)^2-(y2-y5)^2)*((x3-x5)^2+(y3-y5)^2-(x3-x2)^2-(y3-y2)^2)
// p13:=((x6-x2)^2+(y6-y2)^2-(x6-x3)^2-(y6-y3)^2)*((x2-x3)^2+(y2-y3)^2-(x2-x6)^2-(y2-y6)^2)*((x3-x6)^2+(y3-y6)^2-(x3-x2)^2-(y3-y2)^2)
p14:=((x5-x2)^2+(y5-y2)^2-(x5-x4)^2-(y5-y4)^2)*((x2-x4)^2+(y2-y4)^2-(x2-x5)^2-(y2-y5)^2)*((x4-x5)^2+(y4-y5)^2-(x4-x2)^2-(y4-y2)^2)
// p15:=((x6-x2)^2+(y6-y2)^2-(x6-x4)^2-(y6-y4)^2)*((x2-x4)^2+(y2-y4)^2-(x2-x6)^2-(y2-y6)^2)*((x4-x6)^2+(y4-y6)^2-(x4-x2)^2-(y4-y2)^2)
// p16:=((x6-x2)^2+(y6-y2)^2-(x6-x5)^2-(y6-y5)^2)*((x2-x5)^2+(y2-y5)^2-(x2-x6)^2-(y2-y6)^2)*((x5-x6)^2+(y5-y6)^2-(x5-x2)^2-(y5-y2)^2)
p17:=((x5-x3)^2+(y5-y3)^2-(x5-x4)^2-(y5-y4)^2)*((x3-x4)^2+(y3-y4)^2-(x3-x5)^2-(y3-y5)^2)*((x4-x5)^2+(y4-y5)^2-(x4-x3)^2-(y4-y3)^2)
// p18:=((x6-x3)^2+(y6-y3)^2-(x6-x4)^2-(y6-y4)^2)*((x3-x4)^2+(y3-y4)^2-(x3-x6)^2-(y3-y6)^2)*((x4-x6)^2+(y4-y6)^2-(x4-x3)^2-(y4-y3)^2)
// p19:=((x6-x3)^2+(y6-y3)^2-(x6-x5)^2-(y6-y5)^2)*((x3-x5)^2+(y3-y5)^2-(x3-x6)^2-(y3-y6)^2)*((x5-x6)^2+(y5-y6)^2-(x5-x3)^2-(y5-y3)^2)
// p20:=((x6-x4)^2+(y6-y4)^2-(x6-x5)^2-(y6-y5)^2)*((x4-x5)^2+(y4-y5)^2-(x4-x6)^2-(y4-y6)^2)*((x5-x6)^2+(y5-y6)^2-(x5-x4)^2-(y5-y4)^2)
q1:=((x3-x1)^2+(y3-y1)^2)
q2:=((x4-x1)^2+(y4-y1)^2)
q3:=((x5-x1)^2+(y5-y1)^2)
// q4:=((x6-x1)^2+(y6-y1)^2)
q5:=((x3-x2)^2+(y3-y2)^2)
q6:=((x4-x2)^2+(y4-y2)^2)
q7:=((x5-x2)^2+(y5-y2)^2)
// q8:=((x6-x2)^2+(y6-y2)^2)
q9:=((x4-x3)^2+(y4-y3)^2)
q10:=((x5-x3)^2+(y5-y3)^2)
// q11:=((x6-x3)^2+(y6-y3)^2)
q12:=((x5-x4)^2+(y5-y4)^2)
// q13:=((x6-x4)^2+(y6-y4)^2)
// q14:=((x6-x5)^2+(y6-y5)^2)
ei:=eliminate([p1,p2,p3,p5,p6,p8,p11,p12,p14,p17,q1*q2*q3*q5*q6*q7*q9*q10*q12*t-1],[x4,y4,x5,y5,t])
s:=solve(ei,[x3,y3])
\end{Verbatim}
}

The elimination ideal is as follows:
\begin{dmath*}
\langle5 y_{3}^{11}-65 y_{3}^{9}-1680 x_{3}^{4} y_{3}^{3}-5418 x_{3}^{2} y_{3}^{5}+186 y_{3}^{7}+2325 x_{3}^{4} y_{3}+31740 x_{3}^{2} y_{3}^{3}-186 y_{3}^{5}-33540 x_{3}^{2} y_{3}+65 y_{3}^{3}-5 y_{3},
\allowbreak
x_{3} y_{3}^{7}-9 x_{3} y_{3}^{5}+25 x_{3} y_{3}^{3}-20 x_{3} y_{3},
\allowbreak
x_{3}^{3} y_{3}^{4}+2 x_{3} y_{3}^{6}-5 x_{3}^{3} y_{3}^{2}-19 x_{3} y_{3}^{4}+5 x_{3}^{3}+55 x_{3} y_{3}^{2}-45 x_{3},
\allowbreak
x_{3}^{5}+2 x_{3}^{3} y_{3}^{2}+x_{3} y_{3}^{4}-10 x_{3}^{3}-6 x_{3} y_{3}^{2}+9 x_{3}\rangle
\end{dmath*}
The possible solutions for $A_3=(x_3,y_3)$ are
\begin{dmath*}
\left(0,0\right),\left(3,0\right), \left(-3,0\right),\left(1,2\right),\left(-1,2\right),\left(1,-2\right),\left(-1,-2\right),
\allowbreak
\left(\frac{\sqrt{5}+1}{2},\sqrt{\frac{\sqrt{5}+5}{2}}\right),\left(\frac{-\sqrt{5}+3}{2},\sqrt{\frac{\sqrt{5}+5}{2}}\right),\left(\frac{\sqrt{5}+1}{2},-\sqrt{\frac{\sqrt{5}+5}{2}}\right),
\allowbreak
\left(\frac{-\sqrt{5}+3}{2},-\sqrt{\frac{\sqrt{5}+5}{2}}\right),\left(\frac{\sqrt{5}+3}{2},\sqrt{\frac{-\sqrt{5}+5}{2}}\right),\left(\frac{-\sqrt{5}+1}{2},\sqrt{\frac{-\sqrt{5}+5}{2}}\right),
\allowbreak
\left(\frac{\sqrt{5}+3}{2},-\sqrt{\frac{-\sqrt{5}+5}{2}}\right),\left(\frac{-\sqrt{5}+1}{2},-\sqrt{\frac{-\sqrt{5}+5}{2}}\right),\left(\frac{\sqrt{5}-3}{2},\sqrt{\frac{\sqrt{5}+5}{2}}\right),
\allowbreak
\left(\frac{-\sqrt{5}-1}{2},\sqrt{\frac{\sqrt{5}+5}{2}}\right),\left(\frac{\sqrt{5}-3}{2},-\sqrt{\frac{\sqrt{5}+5}{2}}\right),\left(\frac{-\sqrt{5}-1}{2},-\sqrt{\frac{\sqrt{5}+5}{2}}\right),
\allowbreak
\left(\frac{\sqrt{5}-1}{2},\sqrt{\frac{-\sqrt{5}+5}{2}}\right),\left(\frac{-\sqrt{5}-3}{2},\sqrt{\frac{-\sqrt{5}+5}{2}}\right),\left(\frac{\sqrt{5}-1}{2},-\sqrt{\frac{-\sqrt{5}+5}{2}}\right),
\allowbreak
\left(\frac{-\sqrt{5}-3}{2},-\sqrt{\frac{-\sqrt{5}+5}{2}}\right),\left(0,1\right),\left(0,-1\right),\left(0,\sqrt{2 \sqrt{5}+5}\right),\left(0,-\sqrt{2 \sqrt{5}+5}\right),
\allowbreak
\left(0,\sqrt{-2 \sqrt{5}+5}\right),\left(0,-\sqrt{-2 \sqrt{5}+5}\right),\left(0,\sqrt{\frac{2 \sqrt{5}+5}{5}}\right),\left(0,-\sqrt{\frac{2 \sqrt{5}+5}{5}}\right),
\allowbreak
\left(0,\sqrt{\frac{-2 \sqrt{5}+5}{5}}\right),\left(0,-\sqrt{\frac{-2 \sqrt{5}+5}{5}}\right),
\end{dmath*}
33 points. The computation (by using a recent version of Giac, see \url{https://xcas.univ-grenoble-alpes.fr/documentation/EN.html})
takes less than a minute on a personal computer, depending on many factors. Setting $A_1$ and $A_2$ to the above given values
is crucial: by changing $A_1$ to, say, $(0,0)$, the computation may take about an hour.

\subsection{Geometric realization}
\begin{figure}
\centering
\includegraphics[width=0.6\columnwidth]{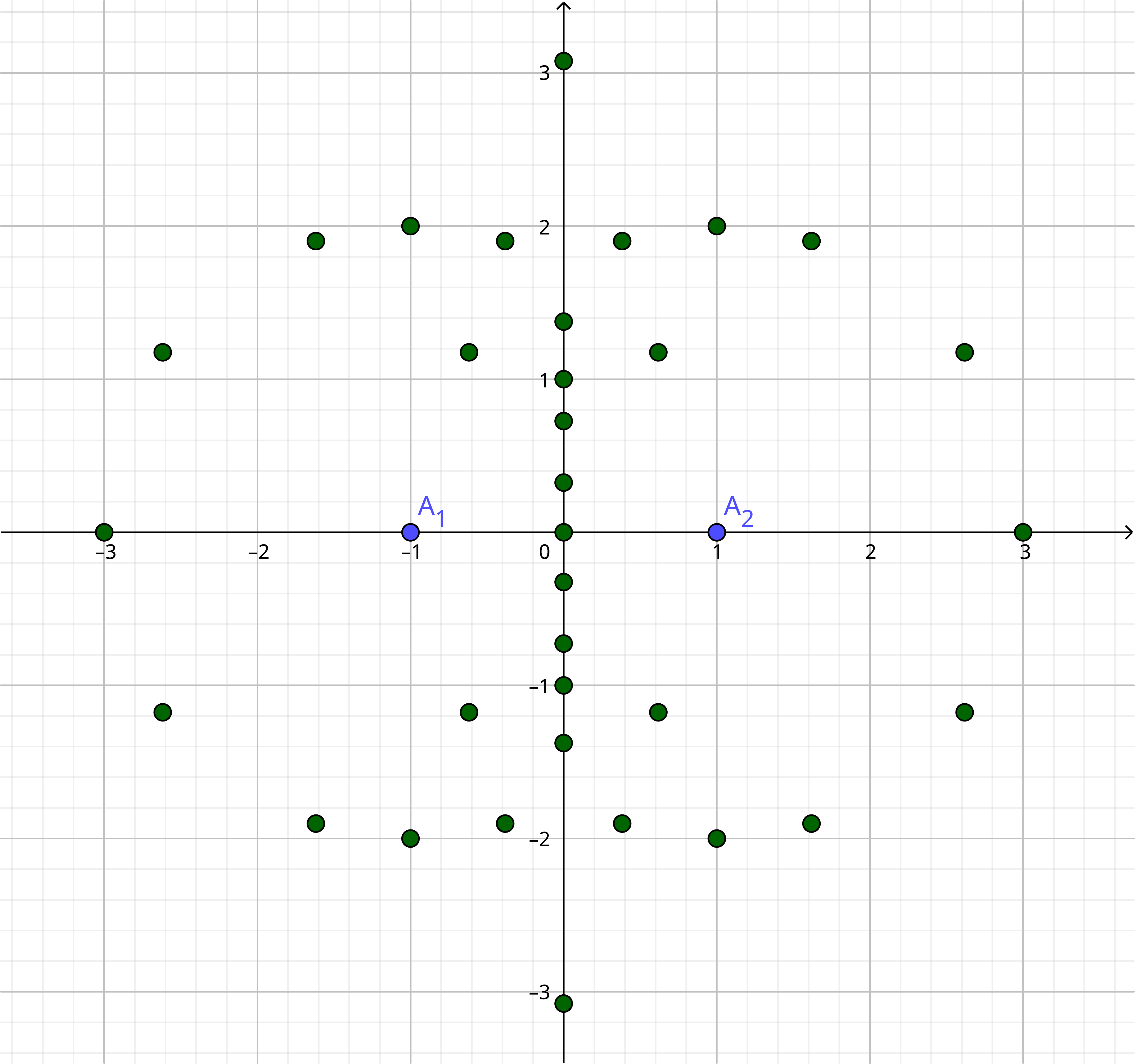}
\caption{A geometric realization of the obtained 33 points.}
\label{fig:33p}
\end{figure}

A geometric realization of the obtained points can be seen in Figure \ref{fig:33p}.

\begin{figure}
\centering
\includegraphics[width=0.6\columnwidth]{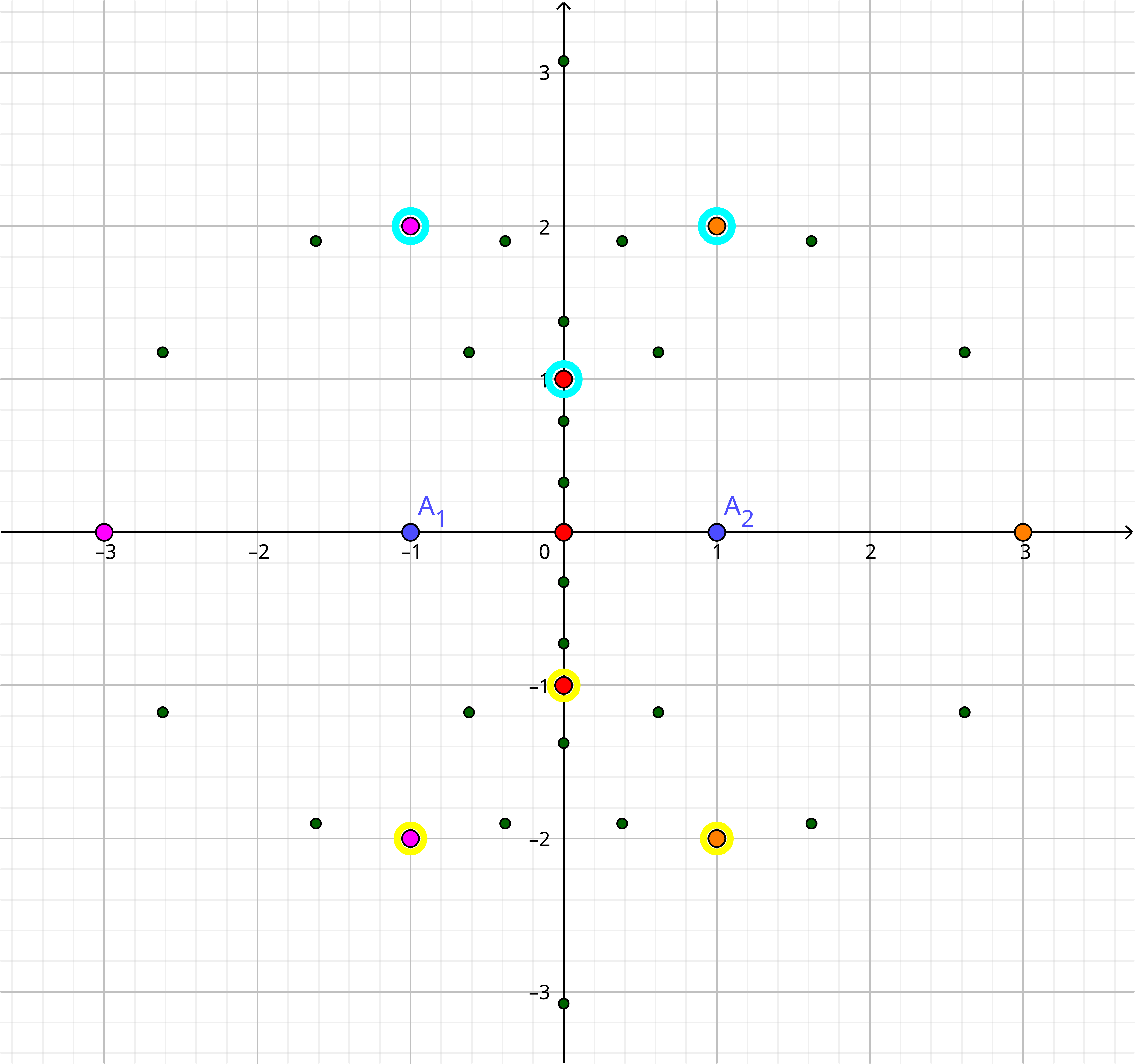}
\caption{Configurations with a \dice{5} shape.}
\label{fig:dice5}
\end{figure}

In Figure \ref{fig:dice5} we identify that $A_1$ and $A_2$ together with 3 additional points may form a set like a \dice{5}.
This affects 9 points that cover 5 sets (some points are covered twice).

\begin{figure}
\centering
\includegraphics[width=0.6\columnwidth]{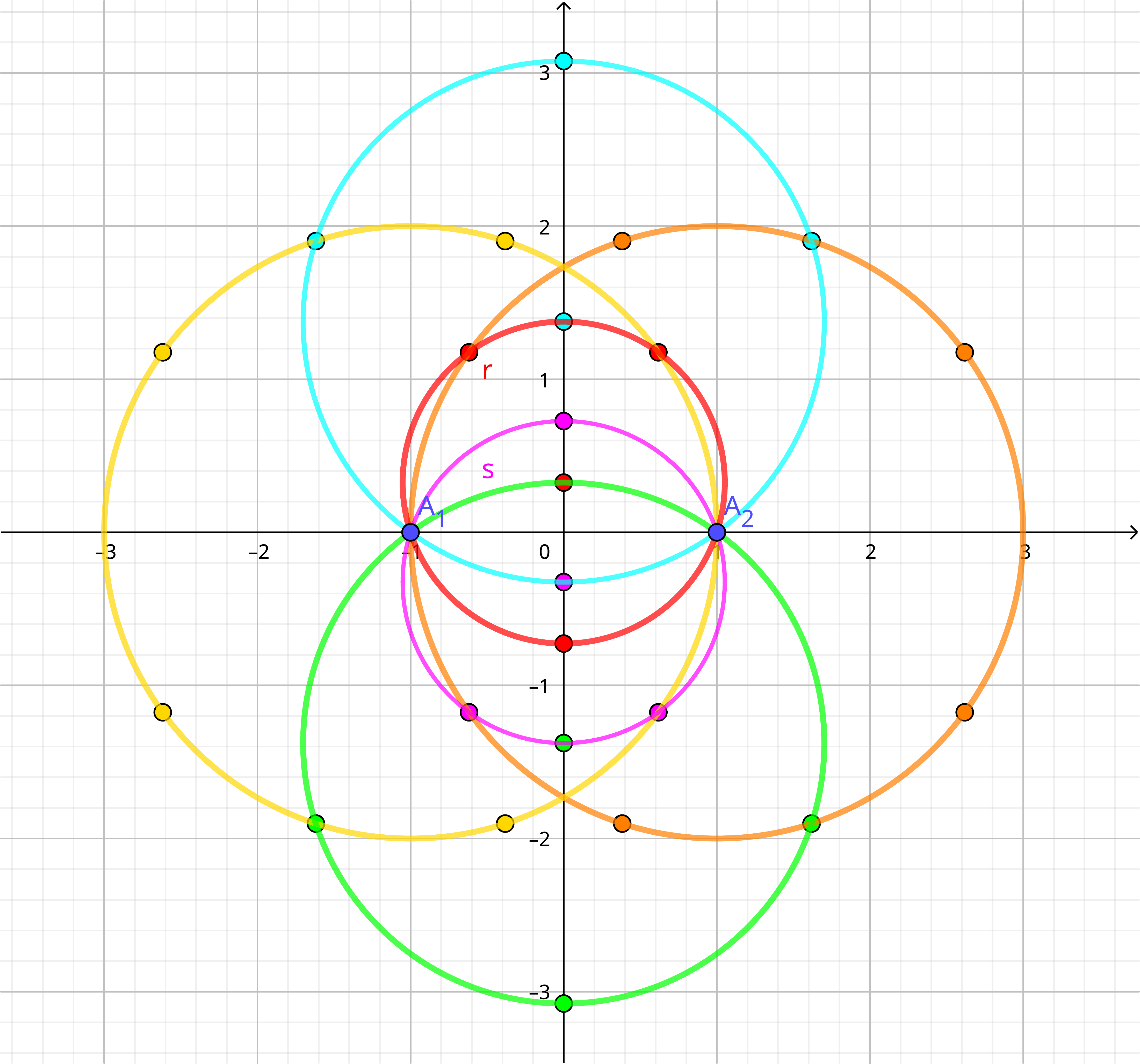}
\caption{Configurations that form a pentagram with its center.}
\label{fig:6sets}
\end{figure}

The remaining 24 points can be classified into 6 sets. Each set consists of 4 points plus $A_1$ and $A_2$: they
appear as 5 vertices of a regular pentagon with the center. The sets are shown in Figure \ref{fig:6sets}.

Now, we have found all possible solutions if $\left|S'\right|=5$: either a \dice{5} shape or a subset of a pentagram with its center.

\section{The solution}
It is easy to see that if a 6th point is added to $S'$, it requires the same assumptions like for the other 5 points.
Therefore, a solution for Erdős's remark must be a 6-point subset of the obtained 33-point set. We already identified
Erdős's conjectured construction among the configurations in Figure \ref{fig:6sets}. The only additional option is
to extend a \dice{5} shape (5 points, two of them are $A_1$ and $A_2$) with another point from the remaining 30 points.
Without loss of generality, it can be assumed that the other 3 points of the \dice{5} shape are $A_3=(0,0)$, $A_4=(0,1)$ and $A_5=(0,-1)$.
Visually it seems quite obvious, but by simple checking it can be seen that this 5-point set cannot be extended
to a 6-point set such that each 3-point subset forms an isosceles triangle. In fact, we can prove this algebraically.
For that we use a simple trick: We put $A_4=(0,0)$, $A_5=(0,1)$ and $A_6=(0,-1)$ (that is, we shift the point indices) and
look for a suitable position for $A_3=(x_3,y_3)$.

Therefore, consider all polynomials $p$ that contain $x_3$ and $y_3$ and drop the other $p$'s. Also, consider all $q$'s that contain
$x_3$ and $y_3$ and leave the other ones. In addition, take all four polynomials in the above-obtained elimination ideal.
Now, compute a new elimination ideal by projecting again to the $(x_3,y_3)$ plane. This must be a simple computation, only in
variables $x_3$, $y_3$ and $t$, because all other variables have already been substituted with integer numbers $-1$, $0$ and $1$.

Hence, by running the following Giac code, we get a quick answer:
{\footnotesize
\begin{Verbatim}[breaklines=true,breakafter=*]
x1:=-1
y1:=0
x2:=1
y2:=0
x4:=0
y4:=0
x5:=0
y5:=1
x6:=0
y6:=-1
p1:=((x3-x1)^2+(y3-y1)^2-(x3-x2)^2-(y3-y2)^2)*((x1-x2)^2+(y1-y2)^2-(x1-x3)^2-(y1-y3)^2)*((x2-x3)^2+(y2-y3)^2-(x2-x1)^2-(y2-y1)^2)
// p2:=((x4-x1)^2+(y4-y1)^2-(x4-x2)^2-(y4-y2)^2)*((x1-x2)^2+(y1-y2)^2-(x1-x4)^2-(y1-y4)^2)*((x2-x4)^2+(y2-y4)^2-(x2-x1)^2-(y2-y1)^2)
// p3:=((x5-x1)^2+(y5-y1)^2-(x5-x2)^2-(y5-y2)^2)*((x1-x2)^2+(y1-y2)^2-(x1-x5)^2-(y1-y5)^2)*((x2-x5)^2+(y2-y5)^2-(x2-x1)^2-(y2-y1)^2)
// p4:=((x6-x1)^2+(y6-y1)^2-(x6-x2)^2-(y6-y2)^2)*((x1-x2)^2+(y1-y2)^2-(x1-x6)^2-(y1-y6)^2)*((x2-x6)^2+(y2-y6)^2-(x2-x1)^2-(y2-y1)^2)
p5:=((x4-x1)^2+(y4-y1)^2-(x4-x3)^2-(y4-y3)^2)*((x1-x3)^2+(y1-y3)^2-(x1-x4)^2-(y1-y4)^2)*((x3-x4)^2+(y3-y4)^2-(x3-x1)^2-(y3-y1)^2)
p6:=((x5-x1)^2+(y5-y1)^2-(x5-x3)^2-(y5-y3)^2)*((x1-x3)^2+(y1-y3)^2-(x1-x5)^2-(y1-y5)^2)*((x3-x5)^2+(y3-y5)^2-(x3-x1)^2-(y3-y1)^2)
p7:=((x6-x1)^2+(y6-y1)^2-(x6-x3)^2-(y6-y3)^2)*((x1-x3)^2+(y1-y3)^2-(x1-x6)^2-(y1-y6)^2)*((x3-x6)^2+(y3-y6)^2-(x3-x1)^2-(y3-y1)^2)
// p8:=((x5-x1)^2+(y5-y1)^2-(x5-x4)^2-(y5-y4)^2)*((x1-x4)^2+(y1-y4)^2-(x1-x5)^2-(y1-y5)^2)*((x4-x5)^2+(y4-y5)^2-(x4-x1)^2-(y4-y1)^2)
// p9:=((x6-x1)^2+(y6-y1)^2-(x6-x4)^2-(y6-y4)^2)*((x1-x4)^2+(y1-y4)^2-(x1-x6)^2-(y1-y6)^2)*((x4-x6)^2+(y4-y6)^2-(x4-x1)^2-(y4-y1)^2)
// p10:=((x6-x1)^2+(y6-y1)^2-(x6-x5)^2-(y6-y5)^2)*((x1-x5)^2+(y1-y5)^2-(x1-x6)^2-(y1-y6)^2)*((x5-x6)^2+(y5-y6)^2-(x5-x1)^2-(y5-y1)^2)
p11:=((x4-x2)^2+(y4-y2)^2-(x4-x3)^2-(y4-y3)^2)*((x2-x3)^2+(y2-y3)^2-(x2-x4)^2-(y2-y4)^2)*((x3-x4)^2+(y3-y4)^2-(x3-x2)^2-(y3-y2)^2)
p12:=((x5-x2)^2+(y5-y2)^2-(x5-x3)^2-(y5-y3)^2)*((x2-x3)^2+(y2-y3)^2-(x2-x5)^2-(y2-y5)^2)*((x3-x5)^2+(y3-y5)^2-(x3-x2)^2-(y3-y2)^2)
p13:=((x6-x2)^2+(y6-y2)^2-(x6-x3)^2-(y6-y3)^2)*((x2-x3)^2+(y2-y3)^2-(x2-x6)^2-(y2-y6)^2)*((x3-x6)^2+(y3-y6)^2-(x3-x2)^2-(y3-y2)^2)
// p14:=((x5-x2)^2+(y5-y2)^2-(x5-x4)^2-(y5-y4)^2)*((x2-x4)^2+(y2-y4)^2-(x2-x5)^2-(y2-y5)^2)*((x4-x5)^2+(y4-y5)^2-(x4-x2)^2-(y4-y2)^2)
// p15:=((x6-x2)^2+(y6-y2)^2-(x6-x4)^2-(y6-y4)^2)*((x2-x4)^2+(y2-y4)^2-(x2-x6)^2-(y2-y6)^2)*((x4-x6)^2+(y4-y6)^2-(x4-x2)^2-(y4-y2)^2)
// p16:=((x6-x2)^2+(y6-y2)^2-(x6-x5)^2-(y6-y5)^2)*((x2-x5)^2+(y2-y5)^2-(x2-x6)^2-(y2-y6)^2)*((x5-x6)^2+(y5-y6)^2-(x5-x2)^2-(y5-y2)^2)
p17:=((x5-x3)^2+(y5-y3)^2-(x5-x4)^2-(y5-y4)^2)*((x3-x4)^2+(y3-y4)^2-(x3-x5)^2-(y3-y5)^2)*((x4-x5)^2+(y4-y5)^2-(x4-x3)^2-(y4-y3)^2)
p18:=((x6-x3)^2+(y6-y3)^2-(x6-x4)^2-(y6-y4)^2)*((x3-x4)^2+(y3-y4)^2-(x3-x6)^2-(y3-y6)^2)*((x4-x6)^2+(y4-y6)^2-(x4-x3)^2-(y4-y3)^2)
p19:=((x6-x3)^2+(y6-y3)^2-(x6-x5)^2-(y6-y5)^2)*((x3-x5)^2+(y3-y5)^2-(x3-x6)^2-(y3-y6)^2)*((x5-x6)^2+(y5-y6)^2-(x5-x3)^2-(y5-y3)^2)
// p20:=((x6-x4)^2+(y6-y4)^2-(x6-x5)^2-(y6-y5)^2)*((x4-x5)^2+(y4-y5)^2-(x4-x6)^2-(y4-y6)^2)*((x5-x6)^2+(y5-y6)^2-(x5-x4)^2-(y5-y4)^2)
q1:=((x3-x1)^2+(y3-y1)^2)
// q2:=((x4-x1)^2+(y4-y1)^2)
// q3:=((x5-x1)^2+(y5-y1)^2)
// q4:=((x6-x1)^2+(y6-y1)^2)
q5:=((x3-x2)^2+(y3-y2)^2)
// q6:=((x4-x2)^2+(y4-y2)^2)
// q7:=((x5-x2)^2+(y5-y2)^2)
// q8:=((x6-x2)^2+(y6-y2)^2)
q9:=((x4-x3)^2+(y4-y3)^2)
q10:=((x5-x3)^2+(y5-y3)^2)
q11:=((x6-x3)^2+(y6-y3)^2)
// q12:=((x5-x4)^2+(y5-y4)^2)
// q13:=((x6-x4)^2+(y6-y4)^2)
// q14:=((x6-x5)^2+(y6-y5)^2)
r1:=5*y3^11-65*y3^9-1680*x3^4*y3^3-5418*x3^2*y3^5+186*y3^7+2325*x3^4*y3+31740*x3^2*y3^3-186*y3^5-33540*x3^2*y3+65*y3^3-5*y3
r2:=x3*y3^7-9*x3*y3^5+25*x3*y3^3-20*x3*y3
r3:=x3^3*y3^4+2*x3*y3^6-5*x3^3*y3^2-19*x3*y3^4+5*x3^3+55*x3*y3^2-45*x3
r4:=x3^5+2*x3^3*y3^2+x3*y3^4-10*x3^3-6*x3*y3^2+9*x3
ei:=eliminate([p1,p5,p6,p7,p11,p12,p13,p17,p18,p19,q1*q5*q9*q10*q11*t-1,r1,r2,r3,r4],[t])
\end{Verbatim}
}
The answer is $\langle1\rangle$, and this means that there is no solution (the system is contradictory), that is, the \dice{5} shape cannot be extended
with a 6th point to fulfill the requirements of Problem E~735. Only the originally stated configuration is possible.

Clearly, by considering the configuration with the regular pentagon and its center, it is impossible to add a 7th point such that for any of its triplets form isosceles triangles. A simple argument why this is impossible: Consider a set $\hat{S}$ with 7 points that satisfies the requirements. Then for any 6-point subsets $S_1\subset\hat{S}$ and $S_2\subset\hat{S}$ we have a regular pentagon with its center (perhaps they are different). Now, $S_1$ and $S_2$ are identical, except for at most one point. This means that at least 5 points of them are identical. At least 4 of these points are vertices of the same regular pentagon, so they are adjacent vertices. Then the remaining 2 points of the 6-point subset must be the 5th vertex of the pentagon and its center, therefore $S_1=S_2$ and $\hat{S}=S_1=S_2$. But in this case $\left|\hat{S}\right|=6<7$.

Therefore, E~735 is solved by means of algebraic geometry and utilizing a computer.

\section{The mysterious remark}

Now we address the remark of \#91. The statement is:
\begin{quote}
Suppose $A\subset\mathbb{R}^2$ has $\left|A\right|=n$ and minimizes the number of distinct distances between points in $A$. Prove that for large $n$ there are at least two (and probably many) such $A$ which are non-similar.
\end{quote}
In the remark, Erdős states that for $n=5$ the regular pentagon is the unique such set. Now, let us assume that there is another 5-point set $B$ such that the number of distinct distances between points is 2. This means that each triplet in $B$ must be isosceles. Since we proved that the only options for $B$ are the \dice{5} shape and a 5-point subset of the union of a regular pentagon and its center, we only have to check these two options:
\begin{enumerate}
\item For the case of the \dice{5} shape, the number of distinct distances is 3: if the side of the square is $1$, then there are distances $\frac{\sqrt2}2$, $1$ and $\sqrt2$.
\item In the other case, we 
\begin{enumerate}
    \item either have the regular polygon without one vertex but the center, or
    \item all 5 vertices of the regular polygon.
\end{enumerate}
The latter is the conjectured case as the only option. The remaining former case is the regular polygon without one vertex but the center, but again, here are 3 distinct distances: if the side of the regular pentagon is $1$, then $1$, $\sqrt{\frac{\sqrt5+5}{10}}$ (between the center and the vertices) and $\frac{\sqrt5+1}2$ (between non-adjacent vertices).
\end{enumerate}
At the end of the day, we proved -- by using a part of the computer proof -- that Erdős's remark in \#91 was correct, and now we have a complete explanation why.

\section*{Acknowledgments}
The author thanks Thomas Bloom for his helpful comments and clarification of some problem settings of Erdős's.


\begin{thebibliography}{9}
\bibitem{Er90}
Erdős, P., Some of my favourite unsolved problems.
A tribute to Paul Erd\H{o}s (1990), 467-478.

\bibitem{Er97}
Erdős, P., Some of my favourite unsolved problems. Math. Japon. (1997), 527-537.

\bibitem{Er93}
Erdős, P., Néhány kedvenc problémám. Polygon 3(2) (1993), 65-67.

\bibitem{ErKe47}
Erdős, P., Kelly, L.M.,  Elementary Problems and Solutions: Solutions: E735. Amer. Math. Monthly (1947), 227-229.

\bibitem{Ra29}
Rabinowitsch, J.L.: Zum {H}ilbertschen {N}ullstellensatz, Math.~Ann., 102(1) (1929), 520.

\bibitem{ReVe99}
Recio, T., V\'elez, M.P.: Automatic discovery of theorems in elementary geometry.
Journal of Automated Reasoning 23 (1999), 63-82.

\end{thebibliography}
\end{document}